\documentclass[proceedings]{JHEP3}

\PrHEP{PrHEP hep2001}                   
\conference{International Europhysics Conference on HEP}                

\usepackage{epsfig}                   
\def\gesim{\lower0.5ex\hbox{$\:\buildrel >\over\sim\:$}}
\def\lesim{\lower0.5ex\hbox{$\:\buildrel <\over\sim\:$}}
\def\xprd#1#2#3{{\it Phys.\ Rev.}\ {{\bf D{#1}} (#2), #3}}

\def\pr#1#2#3{{\it Phys.\ Rep.}\ {{\bf {#1}} (#2), #3}}
\def\plb#1#2#3{{\it Phys.\ Lett.}\ {{\bf B{#1}} (#2), #3}}
\def\npb#1#2#3{{\it Nucl.\ Phys.}\ {{\bf B{#1}} (#2), #3}}
\def\zfp#1#2#3{{\it Z.\ Phys.}\ {{\bf C{#1}} (#2), #3}}
\def\etal{{\it et al.}}

\def\half{\frac12}
\def\lcal{{\cal L}}
\def\up#1{^{(#1)}}
\def\inv#1{\frac1{#1}}
\def\ocal{{\cal O}}
\def\pb{\bar\varphi}
\def\dert#1{{ d #1 \over d t}}
\def\mh{{{m_H}}}
\def\La{{{\Lambda}}}
\def\mt{{{m_t}}}
\def\et{{{\eta}}}
\def\pb{{\bar\varphi}}

\title{Triviality and stability in effective theories}

\author{\speaker{Jos\'e Wudka}                 
        \\ 
        Department of Physics, University of California, Riverside CA 92521-0413, USA\\                                          
        E-mail: \email{jose.wudka@ucr.edu}}                                                

\author{Bohdan Grzadkowski\\                                             
        Institute of Theoretical Physics, Warsaw University, Ho\.za 69, PL-00-681 Warsaw, POLAND\\                                          
        E-mail: \email{bohdan.grzadkowski@fuw.edu.pl}}

\abstract{In this talk we consider the modifications induced by heavy physics
on the triviality and vacuum stability bounds on the Higgs-boson mass. We parameterize
the heavy interactions using an effective Lagrangian and find that
the triviality bound is essentially unaffected for weakly-coupled
heavy physics. In contrast there are significant modifications in
the stability bound that for a light Higgs boson require a scale of
new physics of the order of a few TeV.}

\begin{document}




\vspace{-6in}
\rightline{\parbox{1.5in}{IFT-35-01\\ UCRHEP-T325}}
\vspace{5.5in}

\paragraph{Introduction} The structure of the gauge symmetry breaking
sector of the standard model (SM) remains one of the most interesting
unknowns of this theory. Assuming a scalar sector with a single doublet,
LEP obtained a bound on the Higgs-boson mass of~\cite{higgs_limit}, $ \mh
>113.2 $GeV (mainly using the reaction $ e^+ e^- \to Z H $). For
this particular scalar sector, existing precision measurements can be
used to provide the  upper limit $ \mh < 220 $GeV~\cite{prec_data}.

These results, despite their being {\em very} model dependent, suggest
the presence of light ($\sim 150$GeV) excitations. Should this be
the case, the SM stability~\cite{vacuum_bounds}
and triviality~\cite{triv_bounds}  bounds strongly favor the
appearance of new physics at scales $ \lesim 100 $TeV. Most calculations
of these bounds assume either a specific model, or else assume pure SM
interactions up to a scale $ \Lambda $ where new physics abruptly
``kicks in''. In this talk we review~\cite{gw} the modifications to these 
results when the new physics has a characteristic scale 
$ \Lambda $ significantly above the Fermi scale, yet it does have 
effects at low energies (albeit suppressed by powers of $ 1/ \Lambda $).

\paragraph{New physics effects}
Assuming $ \La $ is much larger than the Fermi scale allows us to 
parameterize the heavy physics effects below $ \La $ using an
effective Lagrangian of the form
\begin{equation}
{\cal L}_{\rm eff} = {1\over \Lambda^2} \sum_i \alpha_i {\cal O}_i +
O(1/\Lambda^3),
\end{equation}
where the $ {\cal O}_i $ are gauge invariant operators constructed using the SM fields,
and the $ \alpha_i $ (calculable if the
underlying theory is known) parameterize the new-physics effects. 

We also assume that the underlying physics is weakly coupled, so that
operators generated at tree level dominate~\cite{aew}, and that 
chiral symmetry is natural~\cite{thooft}. Then the
leading terms in the effective Lagrangian~\cite{bw}
affecting the bounds on $ \mh $ are~\cite{aew}:
\begin{eqnarray}
&&{\ocal_{\phi}} = \inv3 | \phi|^6 \qquad
{\ocal_{\partial\phi}} = \half \left( \partial | \phi |^2 \right)^2 \qquad
{\ocal_{\phi}\up1} = | \phi |^2 \left| D \phi \right|^2 \cr
&&{\ocal_{\phi}\up3} = \left| \phi^\dagger D \phi \right|^2 \qquad
{\ocal_{t\phi}} = | \phi |^2 \left( \bar q \tilde\phi t + \hbox{h.c.} \right) \qquad
{\ocal_{qt}\up1} = \half \left|\bar q t \right|^2,
\nonumber
\end{eqnarray}
where $\phi$ denotes the SM scalar doublet, $q$ the left-handed
top-bottom isodoublet and $t$ the right-handed top isosinglet~\footnote{
The operator $\ocal_{qt}\up1$ affects the stability bound
only through RG mixing and its effects are small;
other similar operators were not included for this reason.}.
We also define $\eta \equiv  \lambda v^2/\La^2 $,
where $v$ denotes the SM vacuum expectation value, $v\sim 246$GeV. The
full Lagrangian we will use is then  
\begin{eqnarray}
&& \lcal_{\rm tree} = 
-\frac14 ( F^2 + B^2) 
+\left| D \phi \right|^2 
+i \bar q \not\!\!D q
+i \bar t \not\!\!D t 
\cr
&& \qquad + f \left( \bar q \tilde\phi t + \hbox{h.c.} \right)
- \lambda \left(|\phi|^2 -v^2/2\right)^2
+ \sum_i\frac{\alpha_i}{\La^2}{\cal O}_i.
\label{lagrangian}
\end{eqnarray}
It is worth noting that in this model one can do 
perturbative calculations; power divergences are relevant for naturality 
arguments while logarithmic divergences determine the renormalization 
group (RG) evolution of the $ \alpha_i $ in the usual manner~\cite{weinberg}.

\begin{figure}[htb]
\psfull
\leavevmode
\epsfig{file=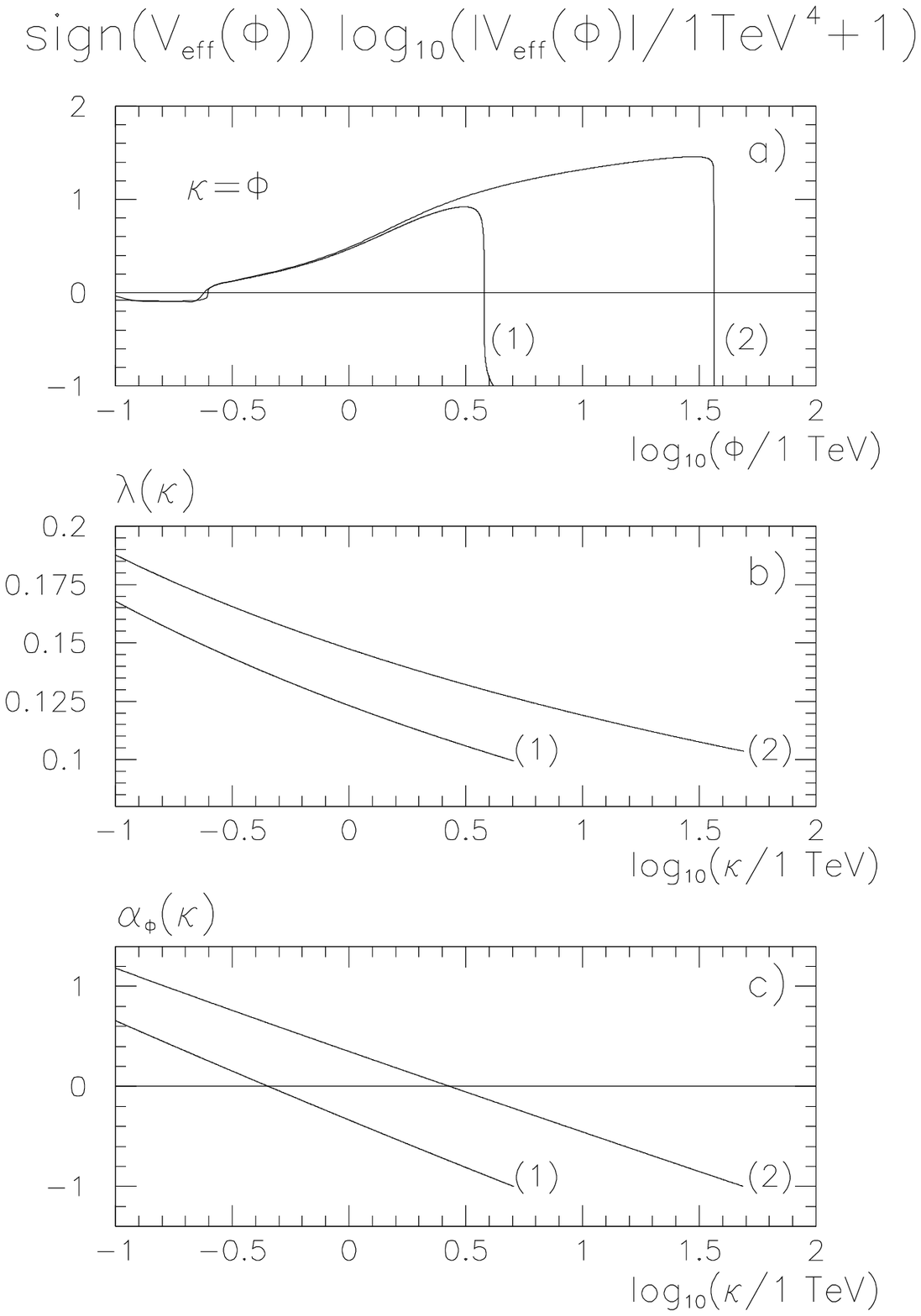,      height=9.2cm, bb=0 0 612 562}
\epsfig{file=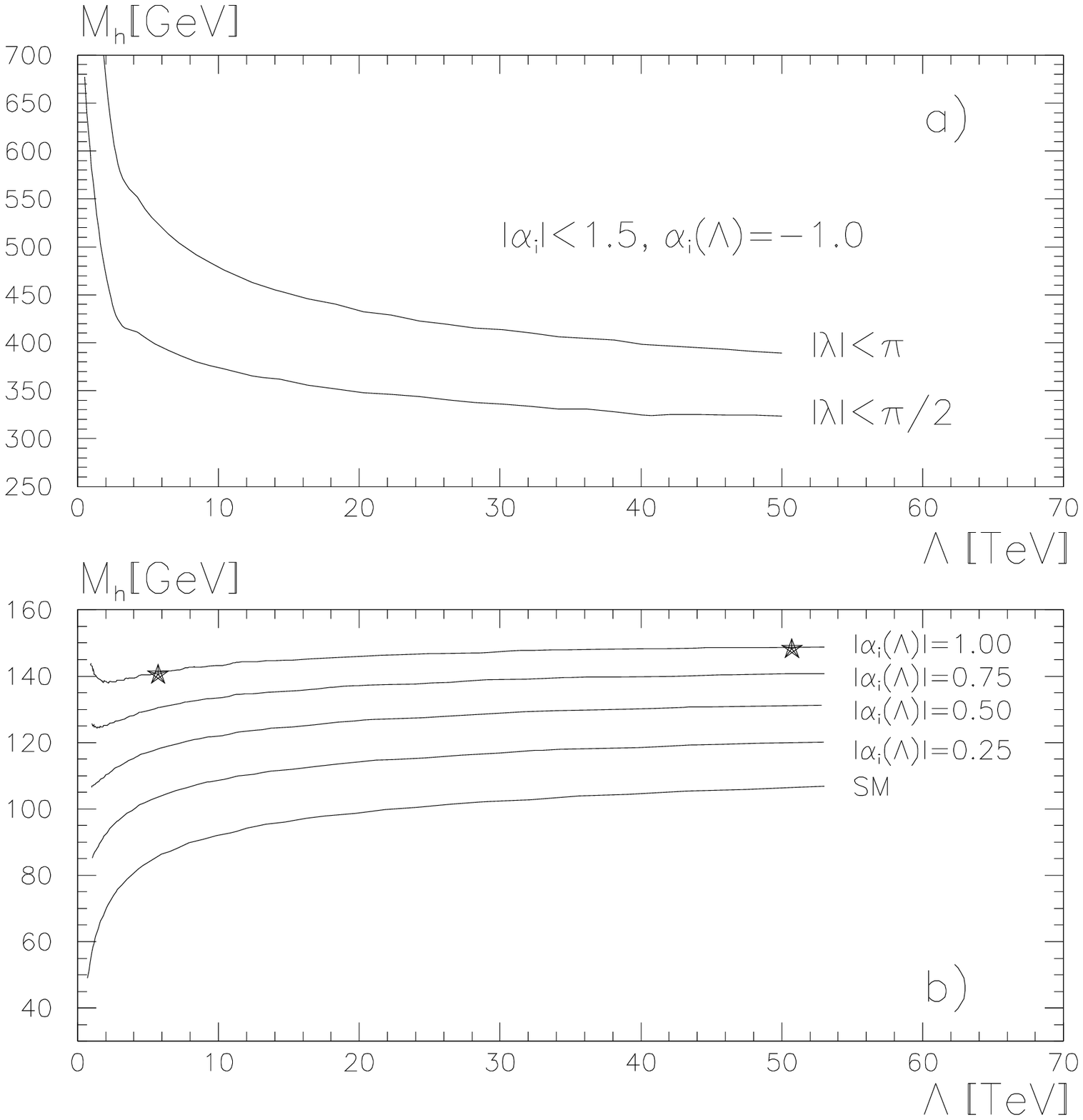, height=9.2cm,width=7cm, bb=100 0 712 562}
\caption{\emph{{\bf Left panel:} (a) $V_{\rm eff}$ at the scale $\kappa=\phi$ 
as a function of the field strength.
The running of $\lambda$ (b) and $\alpha_\phi$ (c) when $ \alpha_i(\La)=-1$,
$\mt = 175 $GeV, for
$ \La =5.1 $TeV, $\mh=140.4 $GeV (curves (1)) and $ \La = 48.9
 $TeV, $\mh = 148.7 $GeV (curves (2)).
{\bf Right panel:} Triviality (a) and stability (b) bounds on $\mh$
for $\mt=175 $GeV.
Stars correspond to solutions (1) and (2).}}
\label{fig:f1}
\end{figure}

\paragraph{Triviality}
We demand that our theory remains perturbative throughout its RG
evolution: 
$|\lambda(\kappa)| < \lambda_{\rm max},~
|\alpha_i(\kappa)| <\alpha_{\rm max} $ for $ \kappa < \Lambda $.
The RG equations to one loop are
\begin{eqnarray}
\dert \lambda &=&
12\lambda^2 -3 f^4 + 6 \lambda f^2 -{3\lambda\over2}\left(3 g^2 + g'{}^2 \right) 
+{3\over16} \left(g'{}^4 +2 g^2 g'{}^2 + 3 g^4\right) 
\cr &&
- 2 \et \left[2 \alpha_\phi + 
\lambda \left( 3 \alpha_{\partial\phi} +4 \bar\alpha + \alpha_\phi\up3\right) \right] 
\cr
\dert \eta &=& 3\eta\left[2\lambda + f^2 - \inv4
\left( 3g^2+ g'{}^2 \right)\right]-2 \et^2 \bar\alpha
\cr
\dert f &=& {9 f^3\over4} -
{f\over2}\left(8 g_s^2 + { 9\over4} g^2 + {17\over12} g'{}^2 \right)
- {f\et\over2}
\left(-6 {\alpha_{t\phi}\over f}+\bar \alpha + 3 \alpha_{qt}\up1 \right) \cr
\dert{\alpha_\phi}&=& 
 9 \alpha_\phi \left(6 \lambda + f^2 \right)
 + 12 \lambda^2 (9\alpha_{\partial\phi}+6 \alpha_\phi\up 1
 +5\alpha_\phi\up 3)
+ 36 \alpha_{t\phi} f^3 
\cr && 
 \hspace{-10pt} - {9\over8}\left[
 2 ( 3 g^2 + g'{}^2) \alpha_\phi+2  \alpha_\phi\up1 g^4 + \left(\alpha_\phi\up1 + 
\alpha_\phi\up3 \right)(g^2 + g'{}^2 )^2 \right] \cr
\dert{\alpha_{\partial\phi}}&=& 2 \lambda
 \left( 6 \alpha_{\partial\phi} - 3\alpha_\phi\up1 +\bar\alpha \right)
 + 6 f \left(f \alpha_{\partial\phi}  - \alpha_{t\phi}\right) \cr
\dert{\alpha_\phi\up1}&=& 2 \lambda 
 \left(\bar\alpha+3\alpha_\phi\up1\right)
 + 6 f \left(f \alpha_\phi\up1 
 -\alpha_{t\phi}\right) \cr
\dert{\alpha_\phi\up3}&=& 6 (\lambda +f^2) \alpha_\phi\up3 \cr
\dert{\alpha_{t\phi}} &=& -3 f (f^2+\lambda) \alpha_{qt}\up1 
+ ({15\over4}f^2-12\lambda) \alpha_{t\phi} 
- {f^3\over2} \left( 
\alpha_{\partial\phi} - \alpha_\phi\up1 + \bar \alpha \right) \cr
\dert{\alpha_{qt}\up1} &=&(3/2) \alpha_{qt}\up1 f^2 ,
\nonumber
\end{eqnarray}
where $ \kappa = M_Z\exp(8 \pi^2 t) $ is the renormalization scale, and
$ \bar\alpha = \alpha_{\partial\phi} +2 
\alpha_\phi\up1 + \alpha_\phi \up3 $. The evolution 
of the gauge couplings
$g$, $g'$ and $ g_s$ (for the strong interactions) is unaffected
by the $\alpha_i$'s.
These equations are solved using the following boundary conditions:
$\alpha_i(\La)= O(1)$ (with various sign choices);
$\langle\phi\rangle = 0.246/\sqrt{2}$TeV (at $\kappa=M_Z$)
and, finally, that the $W,~Z,~t,~H$ masses have their physical
values. Sample plots of the running couplings are given in  Fig.\ref{fig:f1}.

Using the RG solutions and the above condition we find the
triviality bounds plotted in  Fig.~\ref{fig:f1}. These
bounds are indistinguishable from the SM due to our
requirement that the model remains weakly coupled; if this is relaxed
our conclusions need not hold~\cite{chan}.

\paragraph{Stability}
We define the effective potential in terms of the zero-momentum 1PI
$n$-point functions,
$
 V_{\rm eff}(\pb) = - \sum_n  \Gamma\up n(0) \pb^n/n! $
and we require that the SM vacuum is stable, that is
\begin{equation}
V_{\rm eff}(\pb =0.75\La)|_{\kappa=0.75\La} =
V_{\rm eff}(\pb = v_{\rm phys} /\sqrt{2})|_{\kappa=v_{\rm phys} /\sqrt{2}}.
\end{equation}

The calculations were done using the explicit 1-loop expression for $
V_{\rm eff} $,
\begin{eqnarray}
V{\rm eff}(\pb) &=&
- \et \Lambda^2 |\phi|^2 + \lambda |\phi|^4 - 
{\alpha_{\phi} \over 3 \La^2 } | \phi|^6
+ {1 \over 64 \pi^2} \sum_{i=0}^5 c_i R_i^2 [\ln (R_i/\kappa^2)-\nu_i]
+O(1/\La^4),
\nonumber
\end{eqnarray}
where $c_0=-4,~c_1=1,~c_{2,4}=3,~c_3=6,~c_5=-12$, 
$\nu_{0,1,2,5}=3/2,~\nu_{3,4}=5/6$, $ R_0 = \eta\La^2 $, 
\begin{eqnarray}
R_1 &=& \lambda (6 |\pb|^2 -v^2 )\left[1 -
(2 \alpha_{\partial\phi} + \alpha_\phi\up1 + \alpha_\phi\up3 )|\pb|^2/\La^2\right]
- 5 \alpha_\phi |\pb|^4/\Lambda^2 \cr
R_2 &=& \lambda (2 |\pb|^2  -v^2 ) \left[1-(
\alpha_\phi\up1 + \alpha_\phi\up3/3)|\pb|^2/\La^2\right]
 - \alpha_\phi  |\pb|^4/ \Lambda^2 \cr
R_3 &=& (g^2/2)|\pb|^2 \left( 1 +  |\pb|^2 \alpha_\phi\up1/ \Lambda^2  \right) \cr
R_4 &=& [( g^2 +g'{}^2)/2] |\pb|^2 \left( 1 +  |\pb|^2 ( \alpha_\phi\up1 + 
\alpha_\phi\up3) / \Lambda^2  \right) \cr
R_5 &=& f |\pb|^2 \left(f + 2 \alpha_{t \phi} | \pb|^2 / \Lambda^2 \right),\nonumber
\end{eqnarray}
This has the same {\em form} as in the SM, but with {\em modified} $R_i$. 
Note that $ V_{\rm eff}$ is gauge dependent~\cite{Loinaz.Willey} but the 
effects of this gauge dependence are small since  the RG-improved 
tree-level effective potential is gauge-invariant; this leads to a
variation in the 
Higgs-boson mass limit of $ ~ \Delta \mh \lesim 0.5 $GeV~\cite{Quiros}. 
Using the anomalous dimension for the scalar field,
$ \gamma = 3 f^2/2 - 3(3g^2+ g^{\prime 2})/8
- \eta\bar\alpha/2 $, and a careful definition of
$ V_{\rm eff}(0)$~\cite{Ford}, one can verify that $ V_{\rm eff} $ is scale
invariant.

Plots of the effective potential for some representative
values of the parameters and the associated stability
bounds are presented in Fig~\ref{fig:f1}. It is noteworthy 
that in contrast with the triviality bounds the presence of the
effective operators has a significant impact on the stability bounds. For example
for a Higgs-boson mass of $ 115 $GeV, 
\begin{equation}
\begin{array}{ll}
\La \lesim 24\hbox{TeV} \;\;&{\rm for}\;\; |\alpha_i|=0.25\cr
\La \lesim \phantom{2}4\hbox{TeV} \;\;&{\rm for}\;\; |\alpha_i|=0.50\cr
\La \lesim \phantom{2}1\hbox{TeV} \;\;&{\rm for}\;\; |\alpha_i|=0.60\nonumber
\end{array}
\end{equation}

\begin{figure}[htb]
\psfull
\begin{center}
\leavevmode
\epsfig{file=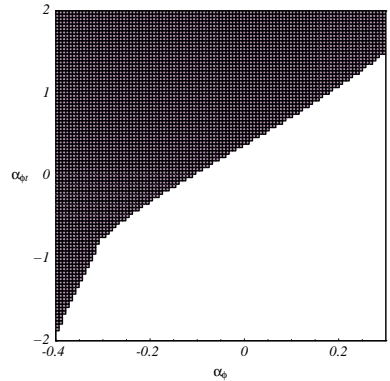,width=5cm}
\end{center}
\caption{\emph{The unshaded region corresponds to the values of
$\alpha_\phi(\La),~\alpha_{\phi t}(\La)$ where the effective
potential has no SM minimum for fields below $0.75 \La $, for any choice
of $0.5 $TeV$< \La <50$TeV.
}}
\label{fig:f2}
\end{figure}

We also find that the main effects on the stability bound are generated by
$\alpha_\phi,~\alpha_{t\phi} $. For example, for $ \alpha_\phi$ large and
positive the potential has no minimum for fields below $0.75 \La $; more precisely,
there is a region in the $\alpha_\phi-\alpha_{t\phi} $, 
given in Fig.\ref{fig:f2},
where the SM vacuum is either absent or unstable for $\pb < 0.75 \Lambda$.

\paragraph{Conclusions}
The SM triviality upper bound remains unmodified for weakly coupled heavy
physics, while the stability bound increases by $\sim50$GeV depending on 
$\La$ and $\alpha_i(\La)$. For $\mh$ close to its lower LEP limit the
constraint on $ \La $ could be
decreased dramatically even for modest values of the $ \alpha_i $
(different values of $ \alpha_i $ correspond to different heavy-physics
theories). These
results complement the ones obtained within specific models~\cite{models}.

Note that, strictly speaking, our expression for $ V_{\rm eff} $
is not valid at points where it changes curvature~\cite{k}.
Still we can make an argument similar to the one above, slightly below the inflection
point $ |\pb| \sim 0.75 \Lambda $;  the resulting bounds are
essentially unchanged due to the precipitous drop of $V_{\rm eff}$ beyond
this point (see Fig.\ref{fig:f1}).

\end{document}